\documentclass[preprint,12pt,authoryear]{elsarticle}

\usepackage{amssymb}
\usepackage{float}
\usepackage{subcaption} 
\usepackage{amsmath}
\usepackage{xcolor}
\usepackage[dvipsnames]{xcolor}
\usepackage{graphicx}
\usepackage{subcaption} 
\usepackage{hyperref}
\graphicspath{{pic/}} 

\journal{Astronomy and Computing}

\begin{document}

\begin{frontmatter}
\title{ArxSP: A Python-Based Modular Application for the Reduction of Digitized Archival Spectra}

\author[FAI]{I.M.~Izmailova}
\author[FAI]{A.Zh.~Umirbayeva}
\author[FAI,KazNU]{M.K.~Khassanov}
\author[FAI]{L.~Aktay}
\author[FAI]{S.A.~Shomshekova}

\affiliation[FAI]{organization={Fesenkov Astrophysical Institute},
            addressline={Observatory 23}, 
            city={Almaty},
            postcode={050020}, 
            country={Kazakhstan}}
\affiliation[KazNU]{%
    organization={al-Farabi Kazakh National University},%
    addressline={al-Farabi Avenue 71},%
    city={Almaty},%
    postcode={050040},%
    country={Kazakhstan}}


\begin{abstract}
We present a methodology for the reduction of archival spectral data together with the description of a newly developed Python-based software package featuring an interactive graphical interface. The work is primarily aimed at processing spectra obtained with electron–optical converters (EOCs), which are characterized by geometric distortions induced by the magnetic field of the registration system. Such data are preserved, in particular, in the archive of the Fesenkov Astrophysical Institute (FAI), which contains about 10,000 photographic plates. These distortions, along with the need to transform the optical density of the photographic material into relative intensity, cannot be corrected by standard astronomical packages such as \texttt{IRAF} and therefore require a dedicated approach. Historically, reductions at FAI were performed using a program written in the \texttt{Microsoft QuickC} language for computing platforms of the 1990s, rendering it incompatible with modern operating systems. The new package is implemented with the \texttt{PyQt5} framework, retaining the logic of the original code while extending its functionality. The implemented algorithms include image rotation and cropping, geometric distortion correction, construction of the characteristic curve linking optical density and intensity, and direct conversion of pixel values in object spectra. The developed software ensures reproducible reduction of archival spectra and provides a cross-platform environment with potential for further extensions.
\end{abstract}
\end{frontmatter}


\section{Introduction}
Archival spectral data play a crucial role in studying the evolutionary changes of astronomical objects, including rare and irregular phenomena such as activity outbursts or short-term variations in emission characteristics \citep{Kondratyeva2025,Shomshekova2022_NAN,Shomshekova2022_NA,Shomshekova2023_ExpAst}. Their unique scientific value stems from the fact that they provide multi-decade temporal baselines that cannot be reproduced with modern observations alone. Spectroscopic monitoring over 30–60 years offers access to slow secular trends, long-term variability cycles, and episodic events that may occur only once in several decades, making such historical datasets indispensable for understanding the physical evolution of active galactic nuclei, planetary nebulae, and other variable astrophysical objects.
The Fesenkov Astrophysical Institute (FAI) maintains an archive of approximately $\sim10\,000$ spectra obtained between 1960 and 1998-a period exceeding three decades of continuous spectroscopic observations. Following the digitization and publication of these data in 2023, part of the archive was integrated into the Kazakhstan Virtual Observatory\footnote{Official webpage: \url{https://vo.fai.kz} (accessed on 26 June 2025).} (KazVO), ensuring long-term preservation and enabling their reuse in modern astrophysical research.

The digitization effort at FAI was inspired by successful international projects such as \texttt{APPLAUSE}~\citep{Enke2024}, \texttt{DFBS}~\citep{Mickaelian2007}, and \texttt{DASCH}~\citep{Grindlay2012}, which highlight the importance of incorporating digitized astronomical archives into modern research practices and virtual observatory infrastructures. Methods of digitization, calibration, and scientific exploitation of plate libraries have been thoroughly described in a number of publications summarizing the experience of leading observatories worldwide \citep{Schechner2016, Jin2007, Nesci2004, Hudec2019, Mironov2007}. These approaches were taken into account in the development of the infrastructure for archival data processing at FAI \citep{Shomshekova2022_NAN, Shomshekova2022_NA, Shomshekova2023_ExpAst}. The raw TIFF images are converted into FIT/FITS format and subsequently undergo a dedicated preprocessing and reduction workflow, described in detail in Section~\ref{sec:prereduction}.

However, the transition from digitization to scientific reduction of spectra involves a number of specific technical challenges. Spectra obtained with three-cascade electron–optical converters (EOCs) are characterized by S-shaped line distortions induced by the magnetic field of the device, and additionally require preliminary geometric corrections such as rotation and cropping of scanned frames. Furthermore, the conversion of optical densities, resulting from emulsion darkening, into relative intensities requires dedicated calibration data. Such operations and distortions are not comprehensively addressed by standard reduction packages such as \texttt{IRAF}~\citep{Tody1986}, \texttt{PyRAF}~\citep{PyRAF}, \texttt{PySpecKit}~\citep{Ginsburg2022}, \texttt{Specutils}~\citep{Specutils2022}, and \texttt{ESO-MIDAS}~\citep{ESO-MIDAS..1996BAAS...28..981.}, which are primarily designed for data with the regular geometry typical of modern CCDs.  

To address these challenges, we have developed the \texttt{ArxSR.py (Archive Spectrum Reduction)} library, intended for comprehensive preprocessing of digitized spectra. It implements the key stages of data preparation, including correction of geometric distortions, filtering of scanning noise and defects, and conversion of optical density into relative intensity based on laboratory calibrations. The library provides classes and methods implementing the essential steps of archival spectrum preprocessing.

In addition, we developed an interactive desktop application \texttt{ARXSP} \texttt{(Archive Spectrum Processing)}. The program serves as a graphical layer on top of the \texttt{ArxSR.py} library, providing researchers with convenient visual access to all stages of preprocessing, including manual markup, calibration, inspection of intermediate results, and export of the final \texttt{.FITS} file. The application is designed both for experienced users and for students and archivists without programming skills, offering a practical tool for handling heterogeneous digitized spectra. 

The remainder of this paper is organized as follows. Section~\ref{sec:software} describes the architecture of the tool, the principles of its modular design, and the implementation of the interactive interface. Section~\ref{sec:prereduction} provides the scientific and technical background of the problem, including the typical distortions of EOC spectra, their specific processing requirements, and the limitations of standard packages. Section~\ref{sec:results} presents the results of applying the program: the volume of published spectra, the number of calibration datasets used, and examples of the resulting FITS files. Finally, Section~\ref{sec:conclusion} summarizes the main findings, discusses the current limitations, and outlines future prospects.

\section{Prereduction} \label{sec:prereduction}

\begin{figure*}[htbp]
    \centering
    \includegraphics[width=0.9\textwidth]{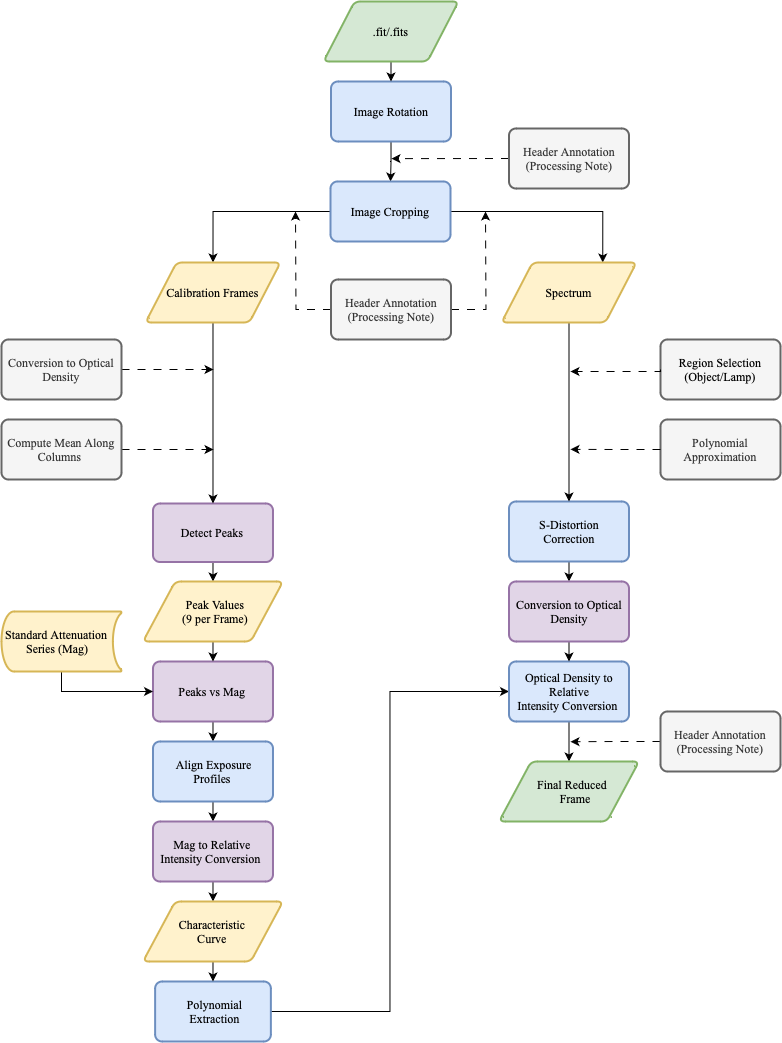}
    \caption{Flowchart describing the preprocessing of archival spectral data. Green -- input and output files; yellow -- intermediate results; blue -- automated processes; purple -- manual processes; grey -- background processes.}
    \label{fig:flwchrt}
\end{figure*}

In developing the preprocessing algorithms for digitized spectral data, we relied on the experience gained with the three-cascade electron–optical converter installed in the spectrograph designed at the Fesenkov Astrophysical Institute in the 1960s. The specific characteristics of this instrument \citep{Denisyuk...2003} required dedicated data processing, for which a specialized software package was implemented in the \texttt{QC} language. A more detailed description of the methods applied to spectra obtained with this instrument is provided in a later work \citep{kondratyeva1985}. The main preprocessing stages implemented in our program are illustrated in the flowchart shown in Fig.~\ref{fig:flwchrt}.

\subsection{Digitization}

The archival photographic spectra were digitized using an Epson Perfection V850 Pro scanner operated in transparent-media mode. The scanner provides stable illumination, uniform response across the field, and sufficient dynamic range to accurately capture the density variations of the photographic emulsion. For all plates, a resolution of 2400 dpi was selected as an optimal compromise between photometric fidelity and file size. Scanning was performed with SilverFast 8, which allows precise control of cropping and the preservation of full tonal depth during acquisition.

Each original photographic plate contains both the astronomical spectrum and the calibration wedge, located in different regions of the same physical plate. Consequently, the full plate is scanned as a single frame, after which the spectral image and the calibration frame are extracted as separate subframes in separate files for further processing.

The raw data are saved in TIFF format and subsequently converted in bulk to 16-bit FIT format using Maxim DL Pro through its batch conversion tool (output parameters: FIT, 16-bit integer, Range mode, Uncompressed). The FIT files are then transformed into standard FITS format within IRAF, where the primary header is created and populated. Additional metadata may be written either directly in IRAF or via a supplementary Python script developed for batch processing.

Before the development of our software, the full preprocessing workflow relied on a heterogeneous combination of tools Maxim DL for format conversion, IRAF for header creation and basic operations, and several standalone Python scripts for specific corrections. This fragmentation complicated the procedure, required manual intervention at multiple stages, and limited reproducibility.

The new \texttt{ArxSR.py} library and the graphical application \texttt{ARXSP} consolidate all these tasks into a single environment: geometric correction, extraction of the spectrum and calibration wedge, filtering, density-to-intensity conversion, metadata generation, and FITS export are now performed within one unified tool. This integration significantly simplifies the processing of large volumes of archival material and ensures consistent application of the same reduction algorithms.

\subsection{Geometric correction }

The spectra were recorded on A-600 photographic film using an electron–optical converter (EOC). Under irradiation, the silver halide in the emulsion underwent photochemical transformations: after development, it was reduced to metallic silver, producing local darkening. This darkening is quantitatively characterized by the optical density:

\begin{equation}
    D = \log \frac{F_0}{F},
\end{equation}

where $F_0$ is the flux of light transmitted through an unexposed region of the film (fog), and $F$ is the flux from the same source transmitted through an exposed region of the same area.

\begin{figure*}[htbp]
  \centering
  \begin{subfigure}{0.99\textwidth}
    \includegraphics[width=0.99\textwidth, trim=0 145 0 0, clip]{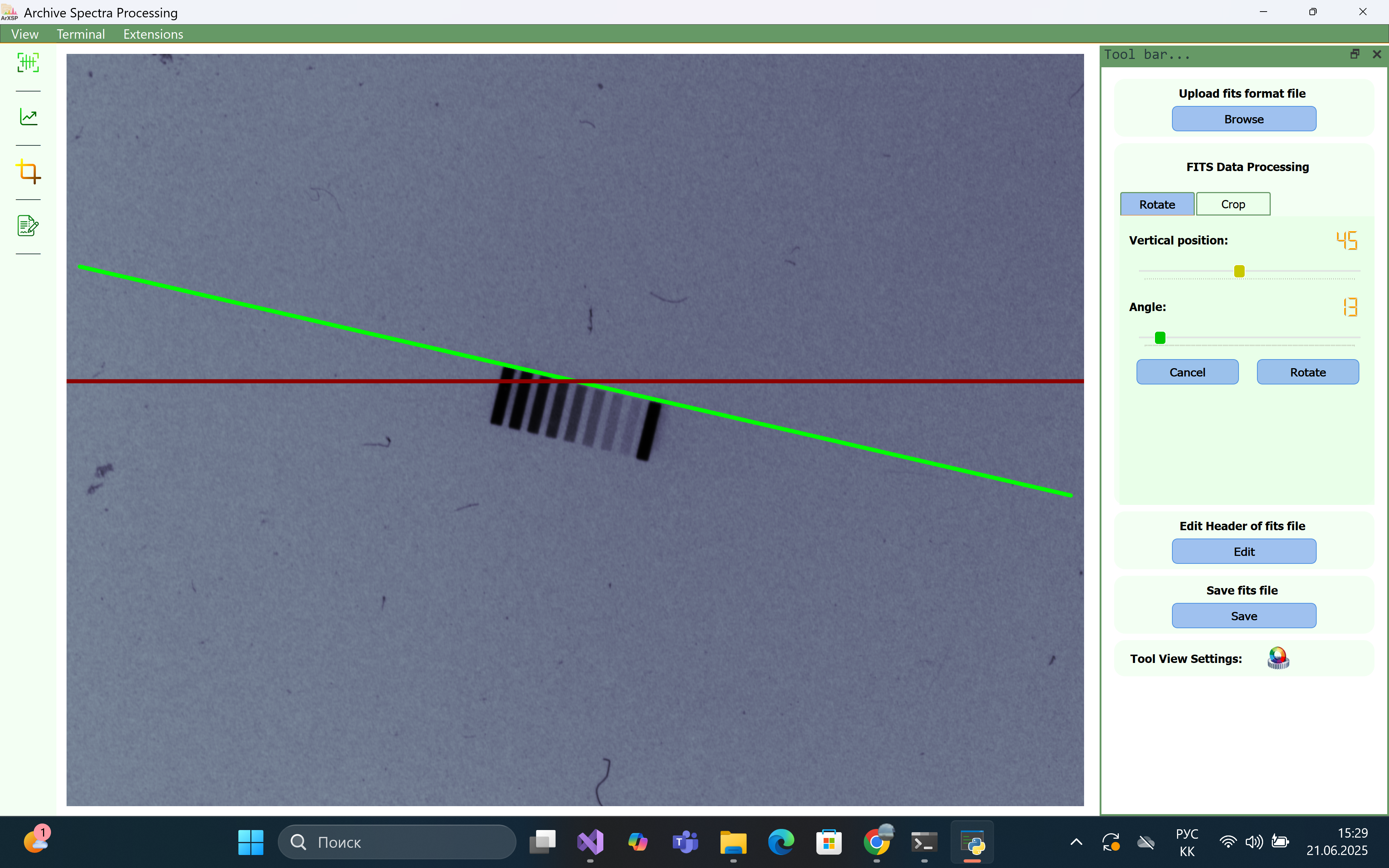}
    \caption*{(a)}
    \label{fig:crop_cal}
  \end{subfigure}
  \hfill
  \begin{subfigure}{0.99\textwidth}
    \includegraphics[width=0.99\textwidth, trim=0 108 0 0, clip]{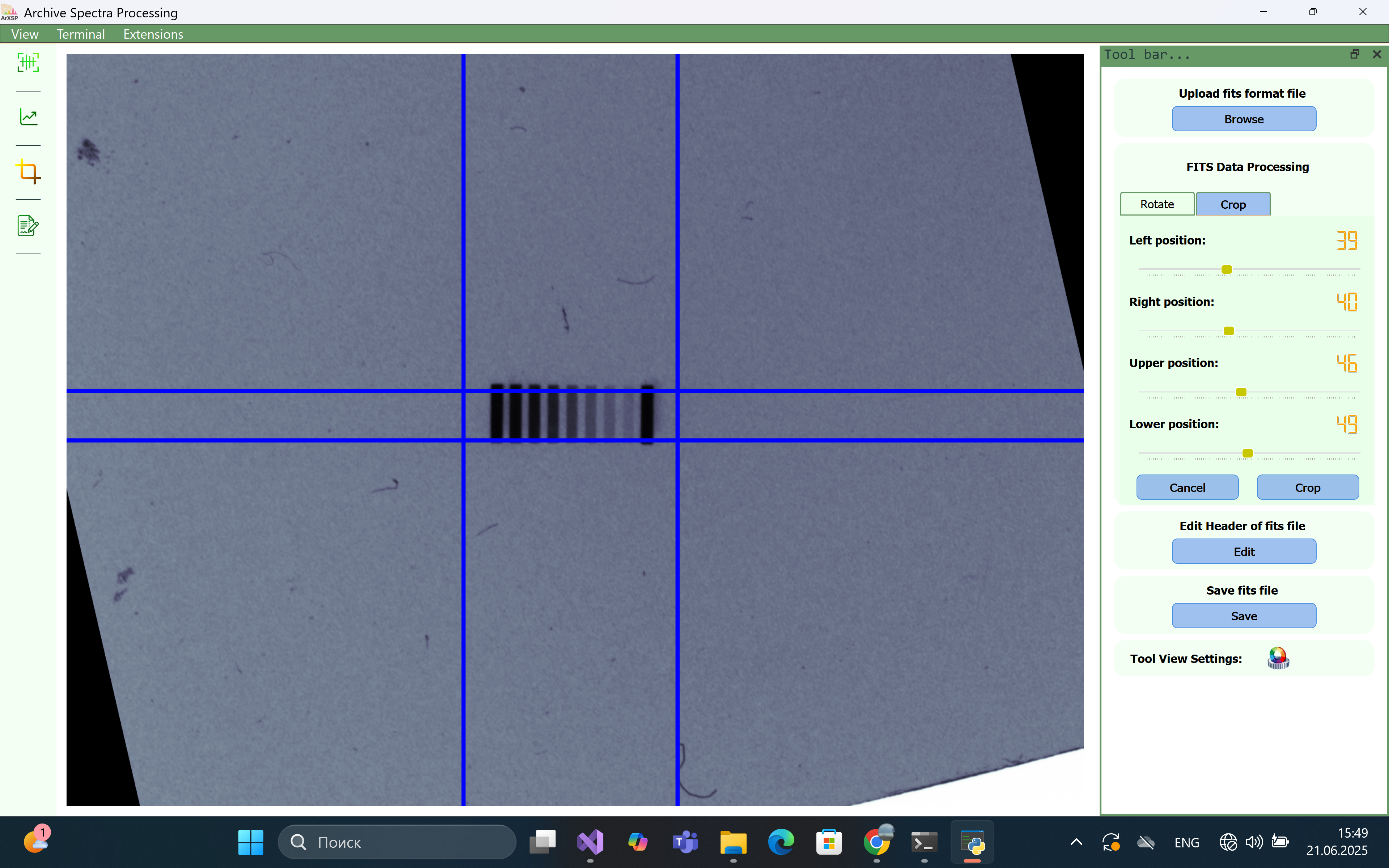}
    \caption*{(b)}
    \label{fig:rot_cal}
  \end{subfigure}

  \caption{Interactive adjustment of preprocessing parameters for a calibration frame: rotation angle (a) and cropping boundaries (b). The central display shows the visualization result with guiding lines to assist in accurate tuning.}
  \label{fig:crop_rot}
\end{figure*}

Determining the relationship between the optical density (darkening) of the negative and the intensity of the incident radiation allows one to construct the characteristic curve. For the correct calculation of optical density, the scanned calibration frame must first be aligned and cropped. In the developed program, these operations are carried out interactively, with user-controlled visual parameter adjustment (Fig.~\ref{fig:crop_rot}).

Frame alignment relative to the horizontal axis is required because calibration scales on the plates were often placed at an angle-both to save space and due to the specifics of the exposure process. Cropping is performed so as to include the step wedge regions at the top and bottom, as well as the adjacent unexposed margins, which helps to eliminate background and external artifacts (scratches, dust, annotations) from further calculations. Preserving these margins is also important for retaining the extreme steps of the wedge, ensuring that they are not cut off by the cropping boundaries and thus remain available for automatic recognition.

\begin{figure}[htbp]
  \centering
  \includegraphics[width=0.99\textwidth, trim=0 108 0 0, clip]{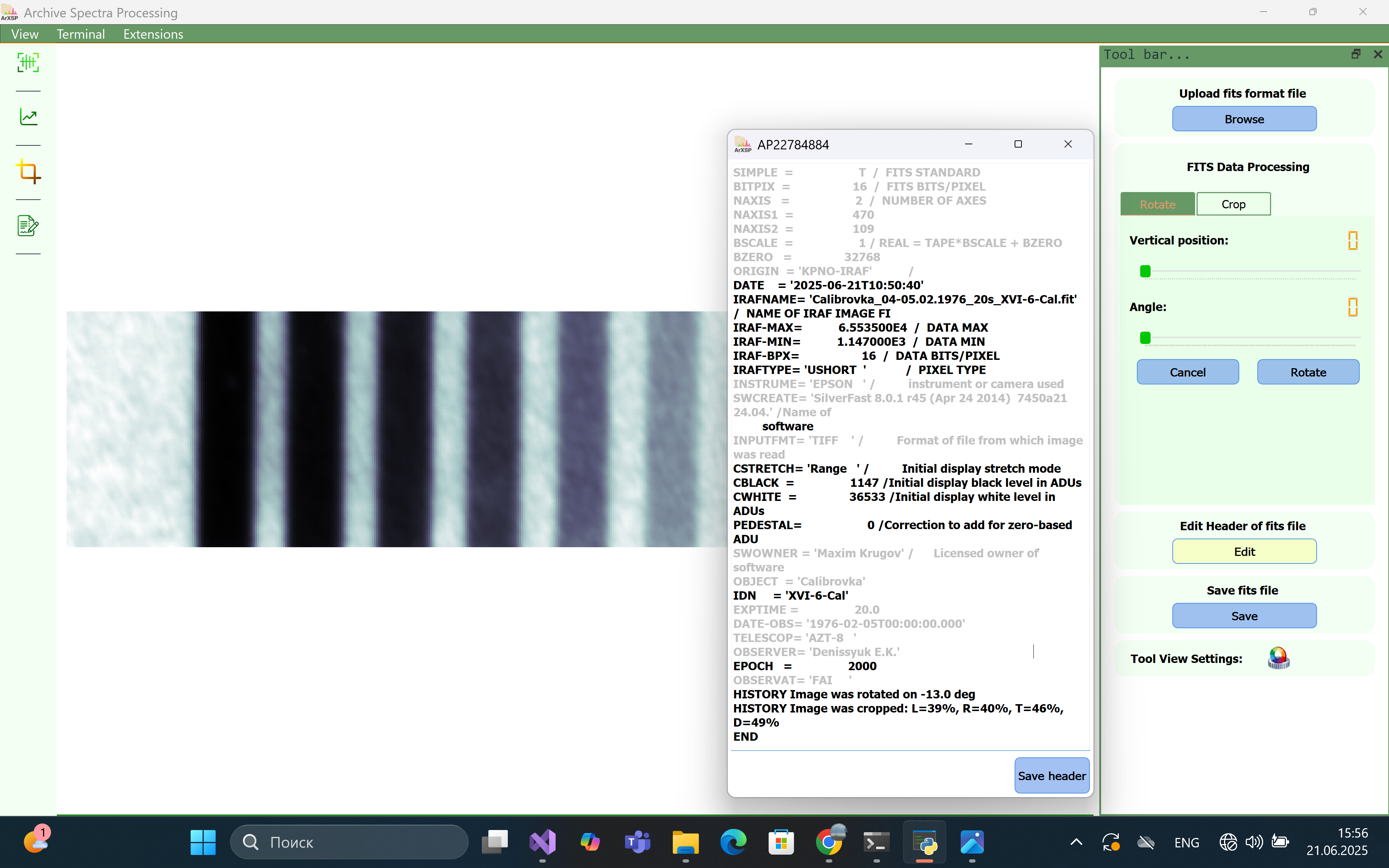}
  \caption{Editing of a \texttt{FITS} file header. The lower part of the window displays \texttt{HISTORY} entries automatically added after each operation. The editing interface allows manual modification of values, while system fields are highlighted in grey and protected from changes.}
  \label{fig:header_edit}
\end{figure}

After each preprocessing operation (e.g., rotation or cropping), a history record is automatically added to the \texttt{FITS} file header. In addition, manual editing is supported with automatic detection of common typographical errors: the user can modify parameter values in the interface together with their corresponding keywords, except for protected system fields, which are highlighted in grey (Fig.~\ref{fig:header_edit}).  

\subsection{Construction of the characteristic curve}

Digitization of the image yields an array of brightness values $x_i$, from which the optical density can be approximated as

\begin{equation}
    X_i = \log \frac{x_{\mathrm{max}}}{x_i} \times 1000,
\end{equation}

where $x_{\mathrm{max}}$ is the maximum brightness (e.g., 65535 for a 16-bit scanner). The logarithmic transformation converts brightness into density, and the factor of 1000 scales the values to a convenient range.  

\begin{figure}[htbp]
  \centering
  \includegraphics[width=0.99\textwidth, trim=0 108 0 0, clip]{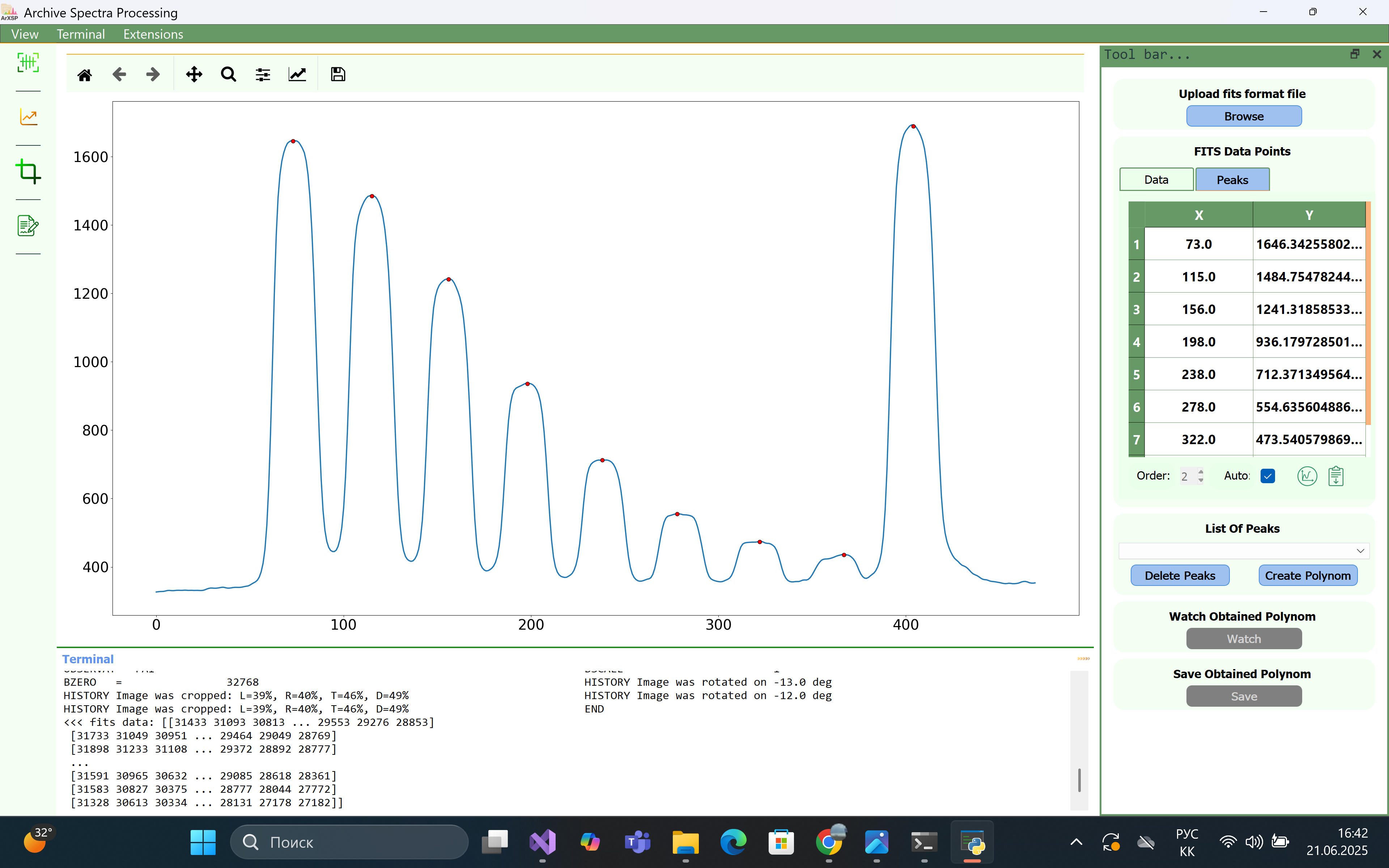}
  \caption{Peak detection corresponding to the calibration wedge steps. Red dots mark local intensity maxima. The right panel displays the parameters for peak detection and visualization, as well as a table with the coordinates of the identified peaks.}
  \label{fig:peaks}
\end{figure}

The mean brightness is then calculated for each image column, producing a one-dimensional exposure profile. In the resulting plot (Fig.~\ref{fig:peaks}), nine distinct peaks are clearly visible, corresponding to the calibration wedge steps. To improve robustness against noise and artifacts, peak values are further averaged over four neighboring points.  

After determining the peak values for constructing the characteristic curve, it is necessary to account for the intensities corresponding to the step wedge attenuator. The attenuator is a quartz plate with platinum stripes of varying thickness, providing different levels of light absorption. During observations, its image was recorded on each photographic plate with different exposures. The step 
intensities are expressed in stellar magnitudes and differ for two calibration 
periods because the attenuator was replaced with a new one on 06 July 1972 
(Table~\ref{table:intensity_step_attenuator}). These “magnitudes” do not represent astronomical photometry; they are an ad-hoc logarithmic transformation (2.5 log I) traditionally used to express the relative attenuation of step-wedge filters. Since steps 1 and 9 have the 
same optical density, only steps 1–8 provide independent information for 
constructing the characteristic curve.

\begin{table*}[ht]
\caption{
Step wedge intensities expressed in stellar magnitudes for two observational 
periods (before and after the replacement of the attenuator on 06 July 1972). 
Steps 1 and 9 have identical optical density.
}
\centering
\begin{tabular}{ccc}
\textbf{\textnumero} & \textbf{Before 06 July 1972} & \textbf{After 06 July 1972} \\
\hline
1 & 0.0 & 0.0  \\
2 & 0.61 & 0.50  \\
3 & 1.10 & 0.97 \\
4 & 1.47 & 1.44 \\
5 & 1.84 & 1.93 \\
6 & 2.25 & 2.43 \\
7 & 2.66 & 2.69 \\
8 & 3.04 & 3.04 \\
9 & 0.0 & 0.0 \\
\hline
\end{tabular}
\label{table:intensity_step_attenuator}
\end{table*}

\begin{figure*}
  \centering

  \begin{subfigure}{0.63\textwidth}
    \includegraphics[width=\textwidth, trim=5 4 5 9, clip]{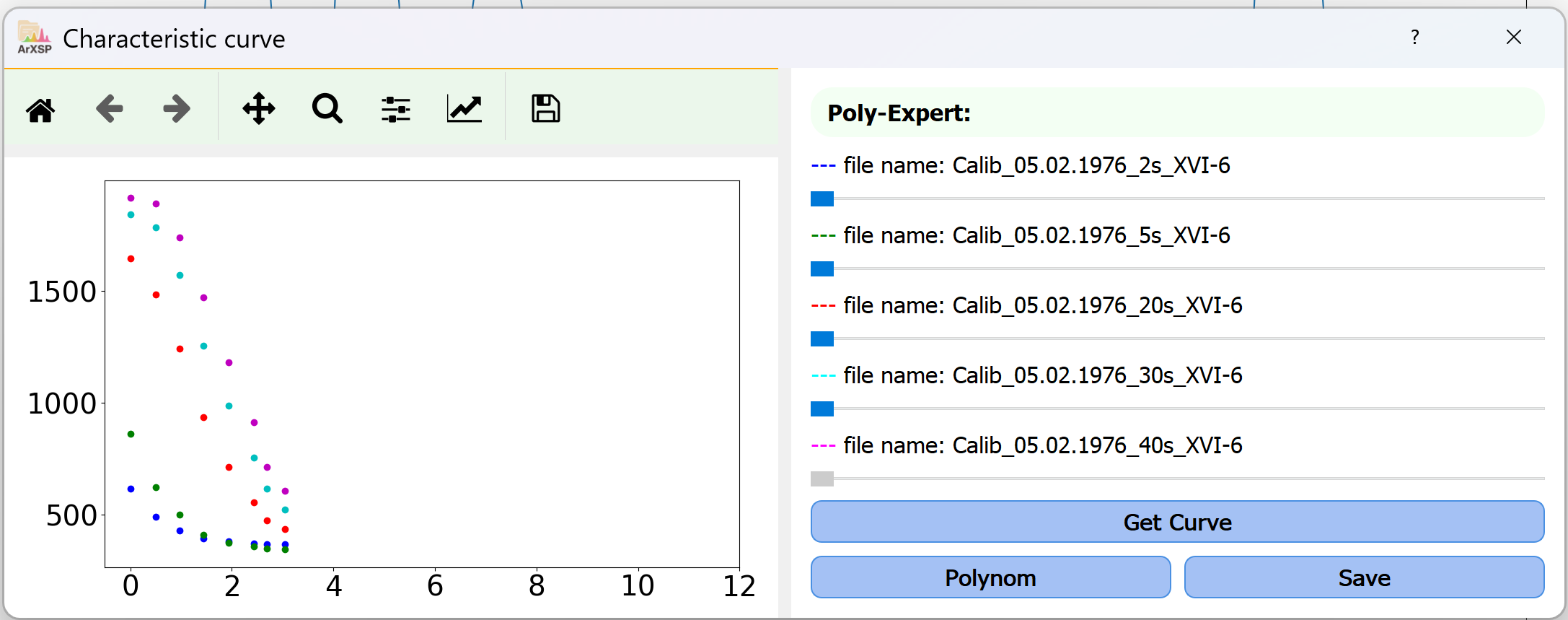}
    \caption{}
    \label{fig:char_curve_peak}
  \end{subfigure} \\[1ex]

  \begin{subfigure}{0.63\textwidth}
    \includegraphics[width=\textwidth, trim=3 3 5 7, clip]{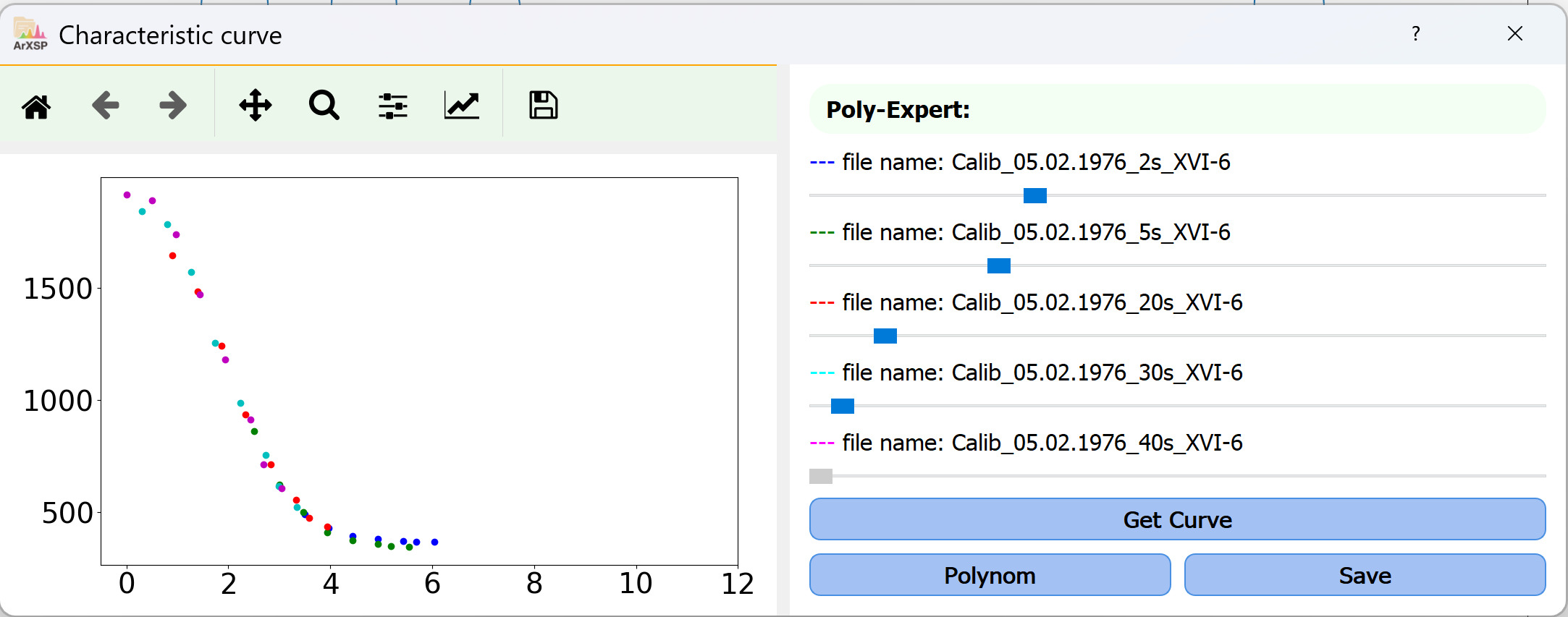}
    \caption{}
    \label{fig:char_curve_peak_aligned}
  \end{subfigure} \\[1ex]

  \begin{subfigure}{0.63\textwidth}
    \includegraphics[width=\textwidth, trim=3 2 0 2, clip]{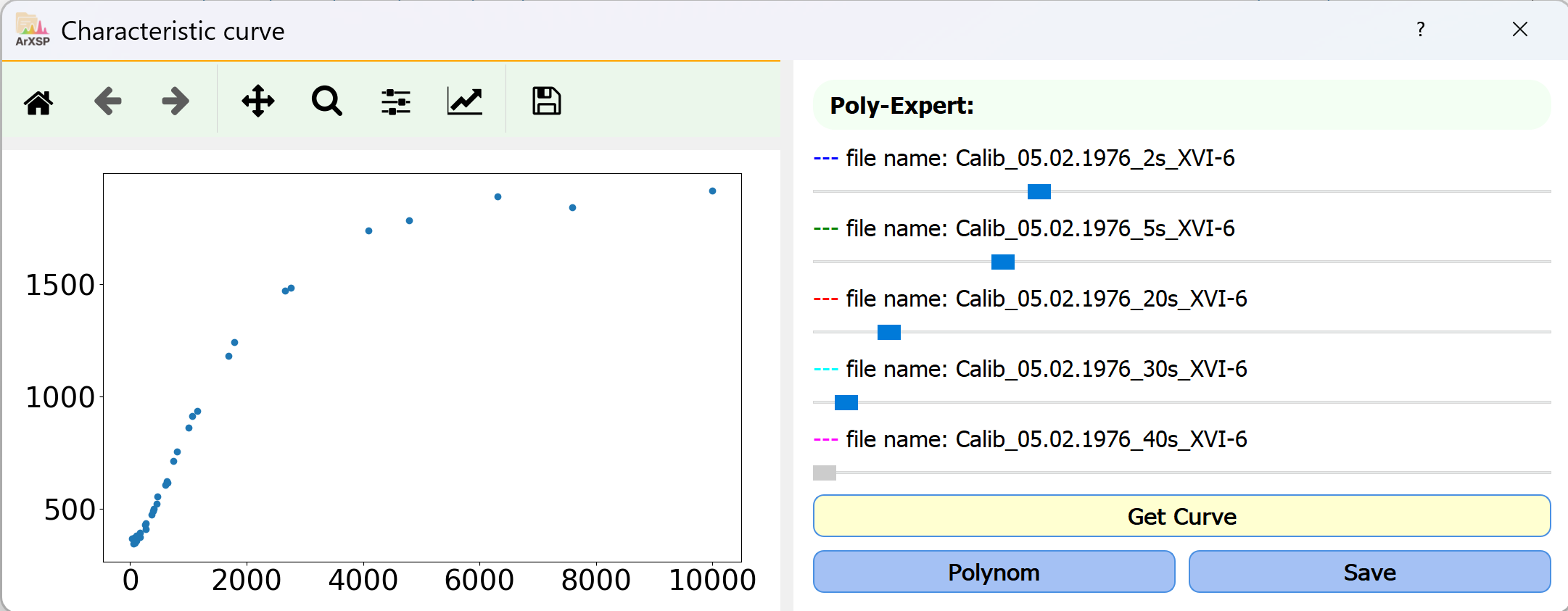}
    \caption{}
    \label{fig:char_curve_intensity}
  \end{subfigure} \\[1ex]

  \begin{subfigure}{0.63\textwidth}
    \includegraphics[width=\textwidth, trim=7 6 0 7, clip]{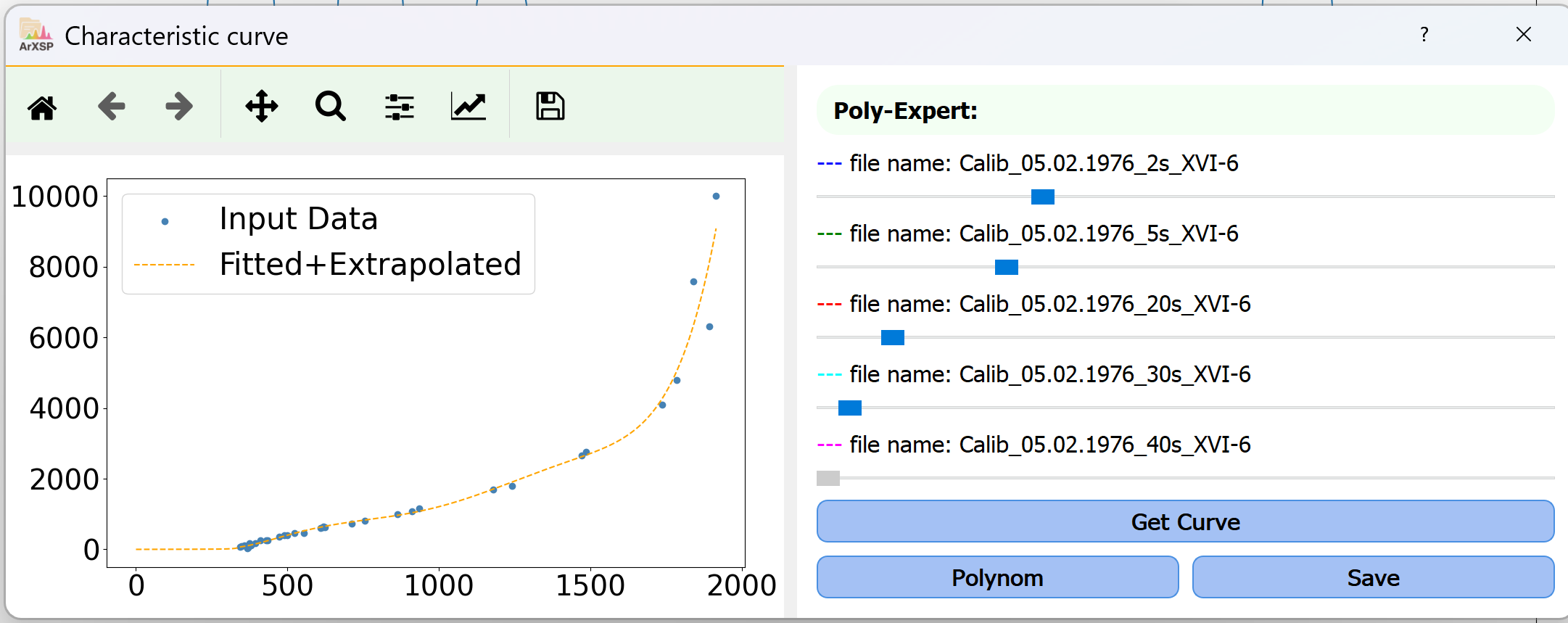}
    \caption{}
    \label{fig:char_curve_polynomial}
  \end{subfigure}

\caption{Stages of constructing the characteristic curve: (a) matching the calibration image peaks to the attenuation values in stellar magnitudes for each step; (b) extending the scale by combining data from different exposures; (c) conversion to a scale of relative intensities; (d) polynomial approximation of the characteristic curve. The right panel of the window shows the controls for manual adjustment of points and for launching the corresponding operations.}
  \label{fig:char_curve}
\end{figure*}

Next, the optical density of the step wedge images is measured. For this purpose, a plot is constructed showing the relationship between the known intensities of the wedge steps and the calculated peak values on the calibration image, which reflect the degree of negative darkening. Figure~\ref{fig:char_curve_peak} presents the data obtained from wedge images with different exposures.

In this plot (Fig.~\ref{fig:char_curve_peak_aligned}), the points show a rather large scatter. The point located at $X = 0$ with the highest degree of darkening corresponds to the densest step obtained at the longest exposure; its ordinate represents the density associated with the maximum intensity of the incident radiation. Measurements acquired at shorter exposures are then shifted along the X-axis toward the position of the longest exposure to bring all datasets into a common reference frame.

In the current version of the software, this alignment is performed manually, following the classical approach traditionally adopted for photographic spectra. Due to the natural variability between measurements obtained at different exposure times, the combined dataset is subsequently approximated by a smooth polynomial function. This procedure suppresses random fluctuations, minimizes user-dependent variability, and yields a stable and reproducible relation that is used to convert photographic darkening into relative intensity.

By aligning and smoothing the data obtained at multiple exposures in this manner, we extend the working scale toward lower relative intensities and construct a combined calibration curve that accurately represents the response of the emulsion across the full dynamic range.

Since the magnitude scale used for the attenuator is not an astronomical 
photometric system but simply a logarithmic representation of optical density 
(2.5 log I), further calculations require conversion to a linear scale of 
relative intensities (Fig.~\ref{fig:char_curve_intensity}). 
The coordinate transformation along the abscissa axis is carried out according 
to the classical relations:

\begin{equation}
    2.5 \log\frac{I_0}{I} = M - M_0,
\end{equation} 
\begin{equation}    
    I = \frac{I_0}{10^{(M - M_0)/2.5}},
\end{equation}

where $I$ is the desired intensity value, $I_0$ is the zero point of the scale (maximum intensity), $M$ is the magnitude of the attenuator step, and $M_0$ is the reference value (in this work $M_0 = 0.0^m$). Since only relative quantities are considered, $I_0$ can be chosen arbitrarily; for convenience we adopt $I_0 = 10\,000$ at $M = 0.0^m$.

The constructed characteristic curve is approximated by high-order polynomials (Fig.~\ref{fig:char_curve_polynomial}). Within the interval covered by the calibration steps, the main polynomial is fitted to minimize the root-mean-square approximation error. Its smoothness and monotonicity are additionally verified, which is essential for ensuring a unique correspondence between optical density and intensity.

Because the exposure density in object spectra may fall below that of the brightest step of the attenuator (e.g., under weak signal conditions or short exposures), it is necessary to extrapolate the characteristic curve toward small $D$ values. For this purpose, the program provides an option to construct an auxiliary polynomial that approximates the function behavior near zero. This polynomial is matched to the main polynomial at a predefined boundary and ensures physically meaningful (monotonically increasing) behavior at low densities.

Thus, the resulting calibration function consists of two polynomial segments -- the main approximation and the extrapolation. The obtained relation is then used to convert darkening values measured in object spectra into relative intensities.

The resulting polynomials can be saved in a local catalog located in the root directory of the program. Each record in the table is accompanied by a unique identifier, the date of the calibration exposures, the coefficients of both the main and extrapolated polynomials, as well as the parameters of their transition. This allows previously constructed curves to be reused in the reduction of archival data.


\subsection{S-distortion correction}
\label{sec:s_distortion}

When scanning spectra obtained with EOCs, a characteristic distortion is the S-shaped bending of spectral lines. This effect is caused by the imperfect geometry of electron focusing and the influence of the magnetic field inside the image intensifier. Such distortions break the linearity of dispersion and can significantly affect the accuracy of subsequent spectral analysis, particularly when extracting a one-dimensional profile.

\begin{figure}[htbp]
  \centering
  \includegraphics[width=0.99\textwidth, trim=0 108 0 0, clip]{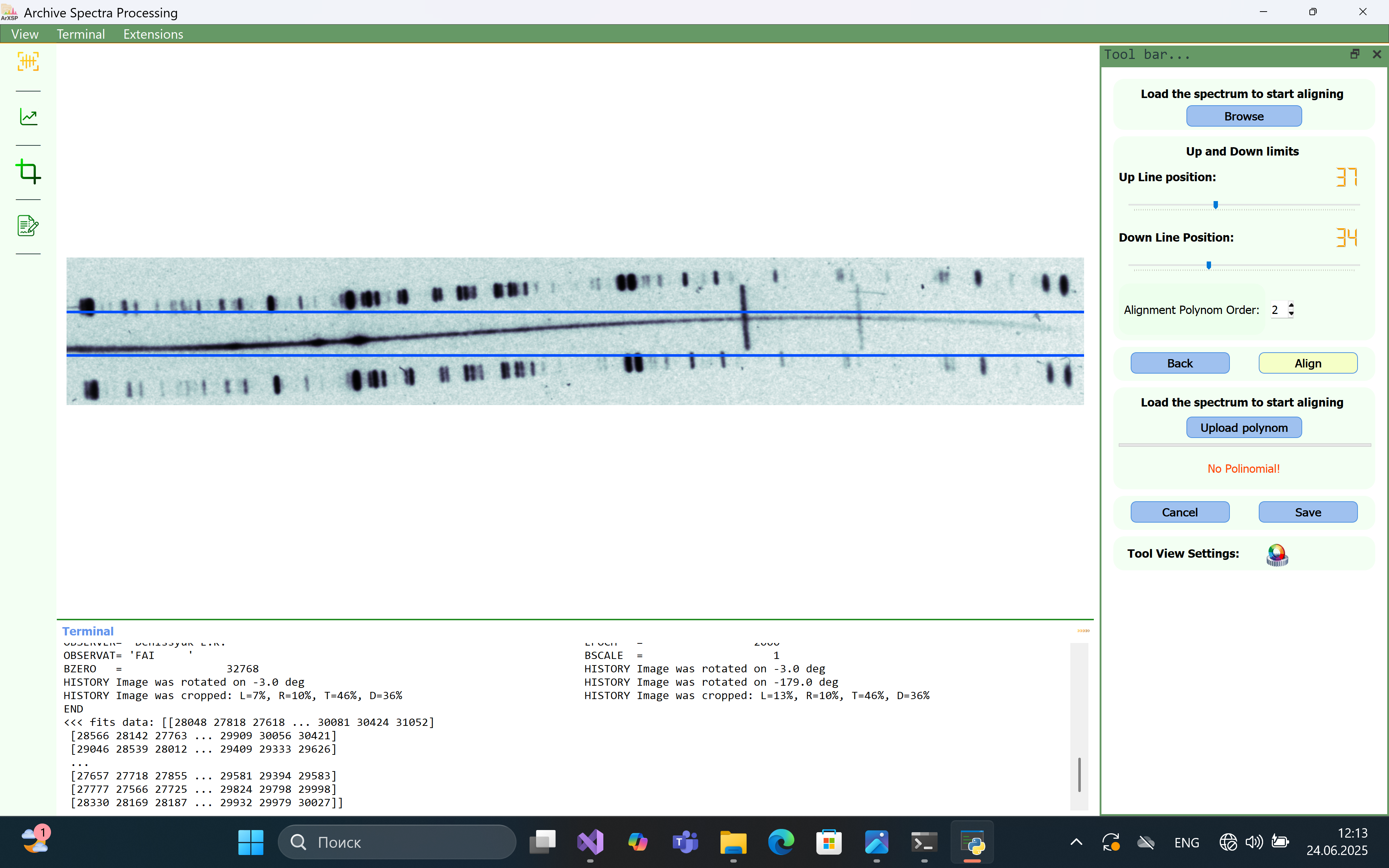}
  \caption{Program interface at the spectrum alignment stage. The calibration spectral image is shown with overlaid aperture lines used to trace the spectrum. The right panel contains controls for line positioning and polynomial order, while the bottom area displays FITS header information.}
  \label{fig:alignment_ui}
\end{figure}

To correct these distortions, the program implements a dedicated alignment stage based on manual spectrum tracing (Fig. \ref{fig:alignment_ui}). During the stage, the user defines two horizontal boundaries that delineate the region containing the spectral track. The program then automatically determines the trace by locating the maximum intensity within this corridor for each column. A polynomial fit is applied to the resulting points to obtain a smooth representation of the spectral path. This procedure minimizes user-dependent variability, since the geometric correction relies on the fitted polynomial rather than on manual point selection.

\subsection{Optical density to relative intensity}
\label{sec:dens_to_intensity}

After geometric distortions have been corrected and the background subtracted, the values in the spectral image still remain in the scale of optical density, determined by the degree of photographic darkening. For subsequent scientific analysis, it is necessary to convert them into a scale of relative intensities. This transformation is carried out using a database of polynomial functions obtained from the calibration data.

\begin{figure}[htbp]
  \centering
  \includegraphics[width=0.99\textwidth, trim=0 108 0 0, clip]{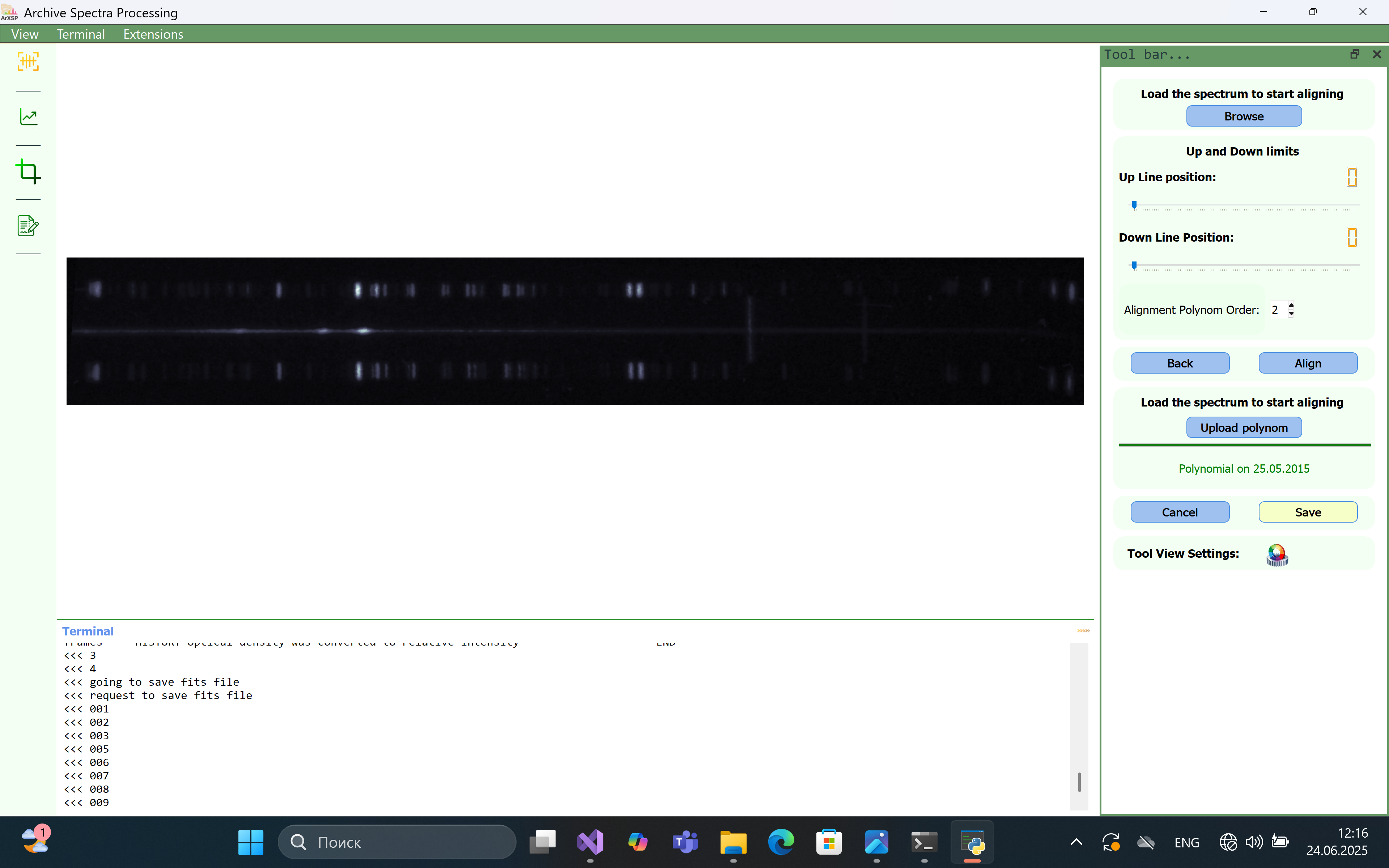}
  \caption{Result of converting spectral image values from optical densities into relative intensities using the polynomial function.} 
  \label{fig:result_poly_applied}
\end{figure}

At this stage, each pixel of the image is transformed according to the function defining the relation between intensity and density. The result is a calibrated spectral image (Fig.~\ref{fig:result_poly_applied}), in which pixel brightness is proportional to the relative intensity of the object’s radiation.

\section{ARXSP software tool} \label{sec:software}

Previously, archival spectra at FAI were processed using a program written in \texttt{Microsoft QuickC}, which was designed for now-obsolete platforms and is incompatible with modern systems. To address this issue, we developed a new application that ensures reproducible preprocessing of archival spectra. The program supports visualization, image rotation, cropping, geometric correction, and the transformation from optical density to intensity, thereby preparing the data for subsequent analysis in \texttt{IRAF}, \texttt{PyRAF}, and other tools. The project architecture provides the flexibility for future extensions, including spectrum extraction, wavelength calibration, and further scientific analysis. The \texttt{ARXSP} program has been tested on \texttt{Windows 11}, \texttt{macOS Ventura (13.5+)} and on \texttt{Ubuntu}-based Linux distributions (20.04 and later).

In addition to the standalone desktop versions, the software has been registered in the catalogue of the Kazakhstan Virtual Observatory (KazVO). Since this entry does not correspond to a typical IVOA data-publication protocol (such as SSAP, SIAP or TAP), the program can be located in the registry by searching for the institute keyword ``\texttt{fai}'', which lists all publicly available FAI services, including the ARXSP software package. The installation packages for \texttt{Linux}, \texttt{macOS}, and \texttt{Windows} are openly accessible on the KazVO website\footnote{\url{https://vo.fai.kz/software.php}}.


\subsection{Programming Language and GUI Framework}

The ARXSP\footnote{\href{https://github.com/ill-i/ArXSP}{\texttt{https://github.com/ill-i/ArXSP}}} program was implemented in \texttt{Python}, chosen for its widespread use in astrophysics and its rich ecosystem of scientific libraries, including \texttt{SciPy}~\citep{Rayhan..2023}, \texttt{Astropy}~\citep{Astropy..AA..2013, Astropy..AJ..2018, Astropy..APJ..2022}, and others~\citep{Harris..Nature..2020, opencv_library, scikit-image, clark2015pillow, matplotlib_Hunter2007, shapely2007}. 

For the development of the graphical user interface, we employed the \texttt{PyQt5} framework \citep{Maulani..DJRCS..2024}, which provides cross-platform compatibility and advanced capabilities for handling data arrays, both of which are critical for the preprocessing of spectral frames.

This technological choice establishes a solid foundation for sustainable development, offering a scalable toolkit potentially applicable to other archival datasets.

\begin{figure}[htbp]
    \centering
    \includegraphics[width=0.99\textwidth]{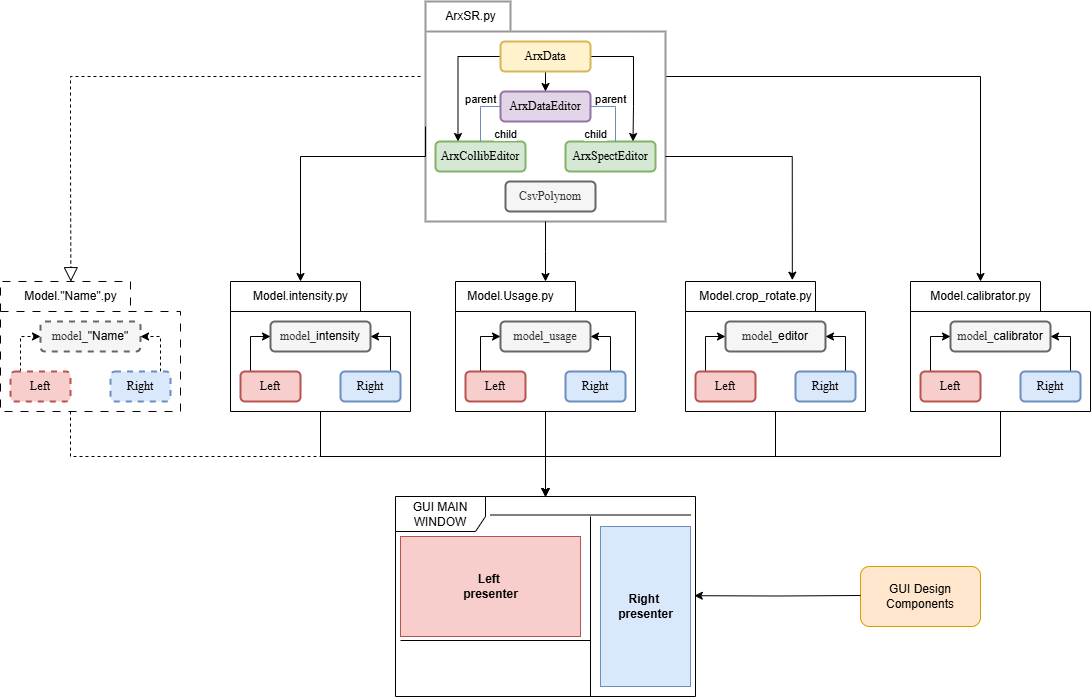}
    \caption{General architecture of the application. The central \texttt{ArxSR.py} module manages the data and editors. Each functional module is connected to the main GUI through two presentation panels.}
    \label{fig:architecture}
\end{figure}

The program is implemented as a modular desktop application with a graphical user interface. Its architecture follows the principle of separating processing logic from visual representation, thereby ensuring extensibility, support for new use cases, and maintainability of the codebase (Fig.~\ref{fig:architecture}).

The \texttt{ARXSP} program has been tested on \texttt{Windows 11}, \texttt{macOS Ventura (Version 13.5)} and later, as well as on \texttt{Ubuntu} distributions version 20.04 and higher.

\subsection{Core library \texttt{ArxSR.py}}

The central component of the \texttt{ARXSP} architecture is the \texttt{ArxSR.py} library, which implements the core logic for processing spectral and calibration data. This module was specifically designed to handle archival files in \texttt{.fit} and \texttt{.fits} formats, providing a convenient data representation structure together with operations for modification, saving, and visualization.

To ensure modularity and extensibility, the library implements a clear class architecture that includes the data model and specialized editors:

\begin{itemize}
    \item \texttt{ArxData} -- a data structure for storing spectrum images, metadata, and calibration information;
    \item \texttt{ArxDataEditor} -- an abstract data editor managing modifications;
    \item \texttt{ArxCollibEditor} and \texttt{ArxSpectEditor} -- derived classes for editing calibration curves and spectra, respectively;
    \item \texttt{CsvPolynom} -- an auxiliary module for handling polynomial calibrations.
\end{itemize}

The \texttt{ArxSR.py} library can be used not only as part of the \texttt{ARXSP} program but also as a standalone module for working with FITS data. It can be installed as an external library in any \texttt{Python} environment by downloading the file from GitHub\footnote{\href{https://github.com/ill-i/ArXSP/blob/main/Pro.AP22784884/ArxSR.py}{\texttt{https://github.com/ill-i/ArXSP/blob/main/Pro.AP22784884/ArxSR.py}}}.

\subsection{Architecture and Processing Modules} \label{sec:modules}

The main shell of the \texttt{ARXSP} program — the \texttt{MainWindow} graphical interface developed with the \texttt{PyQt5} framework — is implemented as a modular system. In the current version, the architecture comprises four core processing modules, each responsible for a distinct stage of spectral data reduction:
\begin{itemize}
    \item \texttt{model.Usage.py} -- introductory module providing usage templates, licensing agreement, and initial user interaction;
    \item \texttt{model.crop\_rotater.py} -- geometric correction of spectra, including cropping and rotation of scanned images;
    \item \texttt{model.calibrator.py} -- construction and analysis of the characteristic curve from calibration frames, enabling the conversion of optical density to relative intensity;
    \item \texttt{model.intensity.py} -- correction of S-shaped distortions and aperture alignment of spectra.
\end{itemize}

This modular architecture localizes specific processing tasks within dedicated components, thereby simplifying maintenance and facilitating further extension of the software package.

Each module in the \texttt{ARXSP} architecture is implemented as a separate directory with the \texttt{model.} prefix, and must contain three key components: \texttt{Model\_``Name''}, \texttt{Left}, and \texttt{Right}:

\begin{itemize}
    \item \texttt{Left} -- a class responsible for the visual representation of data (e.g., displaying an image, plot, or table);
    
    \item \texttt{Right} -- a class defining the control interface: buttons, sliders, input fields, etc.;
    
    \item \texttt{Model\_``Name''} -- the main controller class that implements the business logic of the module, establishes connections between \texttt{Left} and \texttt{Right}, and defines the overall behavior model.
\end{itemize}

Such an architecture enables independent development and integration of modules. The graphical shell automatically detects all modules located in the corresponding subdirectories and dynamically loads them at program startup. This enables flexible extension of the \texttt{ARXSP} functionality by adding new processing stages without the need to modify the existing code. Dedicated methods for automatic registration and the display of control buttons in the interface are provided.

The architecture of the main window and the relationships between its components are shown in Fig.~\ref{fig:architecture}, where the main elements of the \texttt{MainWindow}, as well as the interaction of the toolbar, modules, and editors, are schematically illustrated. This diagram emphasizes the modular principle of the interface design and the mechanism for connecting new components.

For user convenience, installation packages have been prepared for the three operating systems mentioned above. The source code of the program, including the \texttt{ArxSR.py} library, is publicly available on \texttt{GitHub}\footnote{\href{https://github.com/ill-i/ArxSP}{\texttt{https://github.com/ill-i/ArxSP}}} and may be freely used, modified, and integrated into external projects.

\section{Results} \label{sec:results}

\begin{figure*}[htbp]
  \centering
  \begin{subfigure}{0.99\textwidth}
    \includegraphics[width=0.99\textwidth, trim=0 0 0 0, clip]{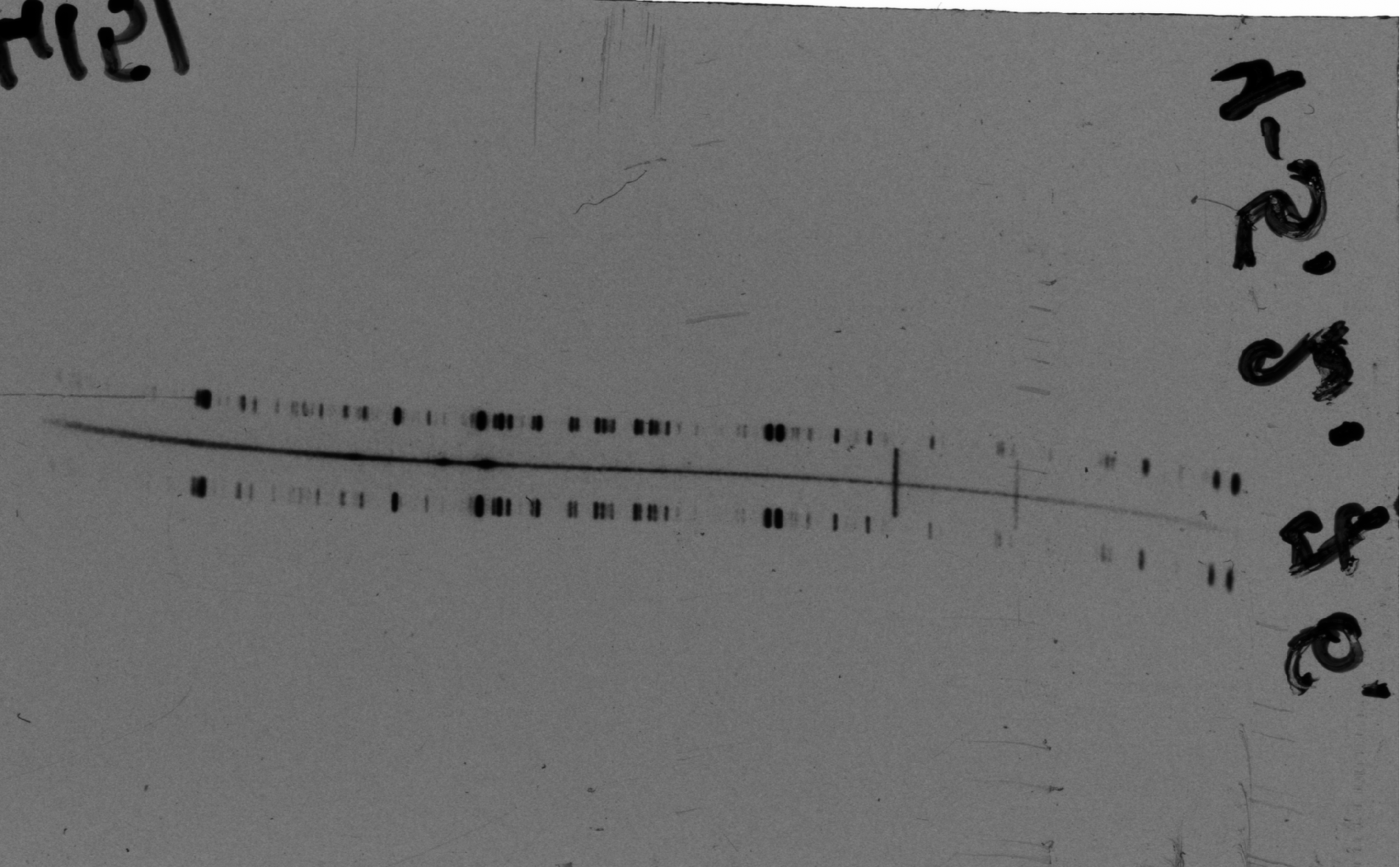}
    \caption{}
    \label{fig:spectra_original}
  \end{subfigure} \\[1ex]
  \begin{subfigure}{0.99\textwidth}
    \includegraphics[width=0.99\textwidth, trim=0 0 0 0, clip]{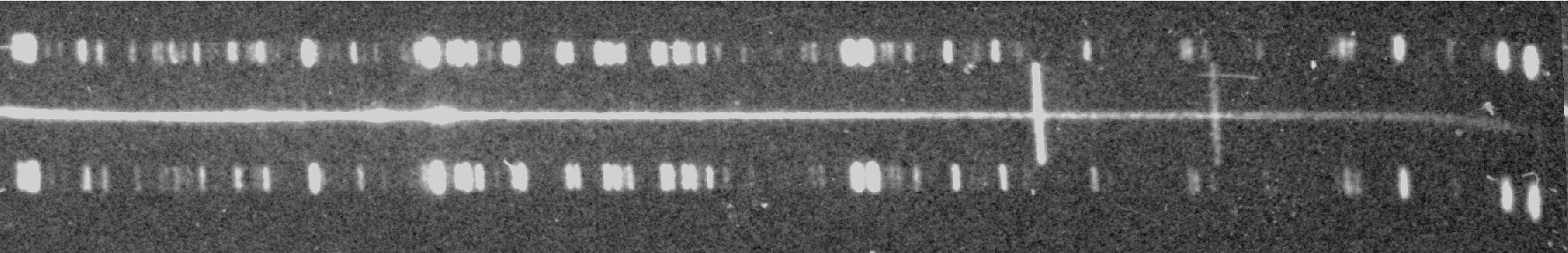}
    \caption{}
    \label{fig:spectra_reducted}
  \end{subfigure}
  \caption{Example of the NGC~4151 spectrum before (a) and after (b) reduction in \texttt{ARXSP}. The correction of S-shaped distortion, aperture alignment, and improved readability of spectral lines are clearly visible.}
  \label{fig:conclusion_example}
\end{figure*}

The efficiency of the developed algorithms is illustrated using the spectrum of the active galactic nucleus NGC~4151 (Fig.~\ref{fig:conclusion_example}). The state of the data before and after reduction with \texttt{ARXSP} is shown. The application of the full set of procedures enabled image cropping, removal of S-shaped distortion, aperture alignment, and conversion of optical density values into relative intensities, which significantly improves the readability of the lines and ensures the correctness of subsequent analysis.  

\begin{figure}[htbp]
  \centering
  \includegraphics[width=\textwidth, trim=0 1 0 42, clip]{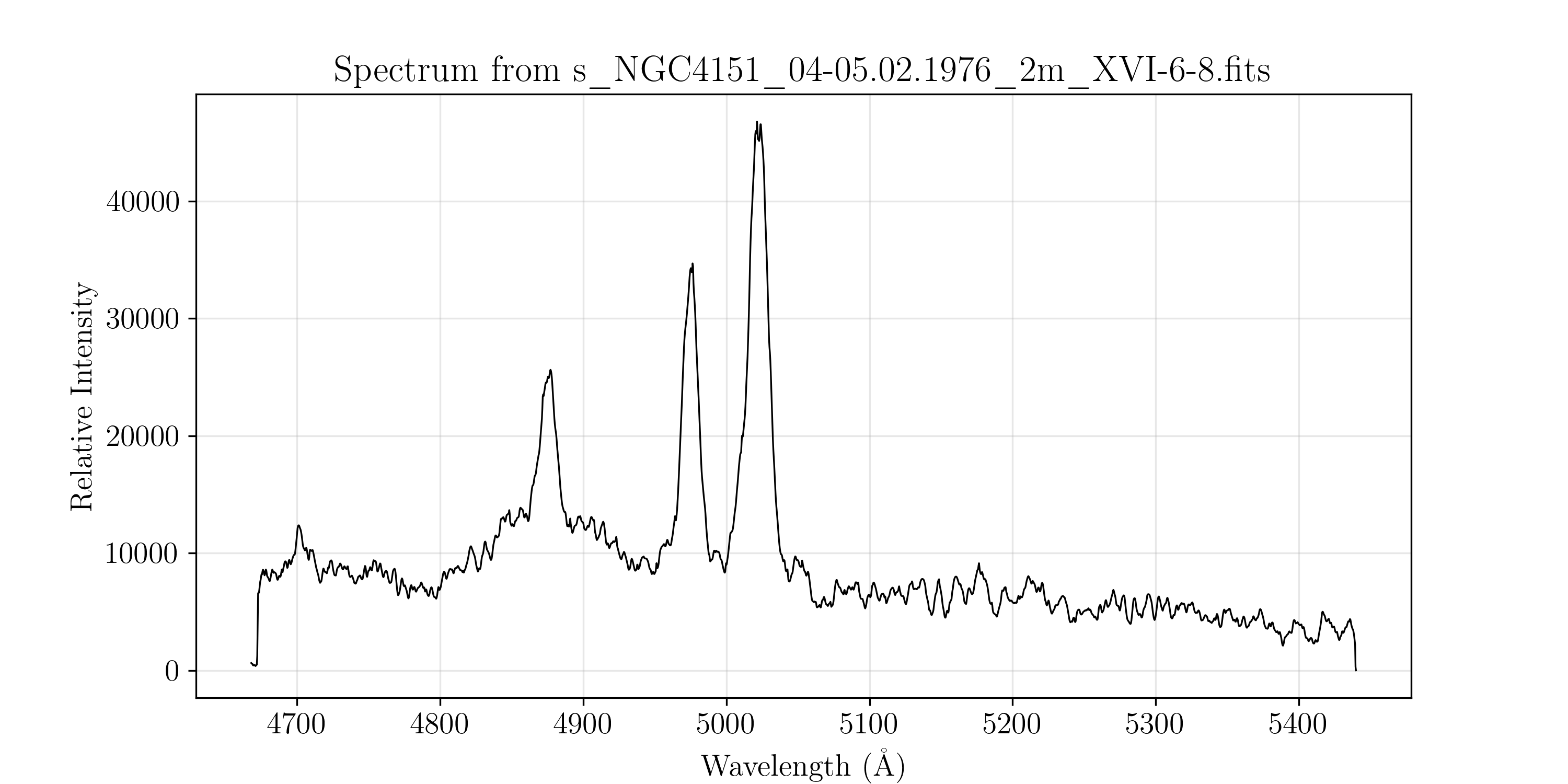}
  \caption{Final spectrum of the active galactic nucleus NGC~4151 after reduction in \texttt{ARXSP} and subsequent processing in \texttt{IRAF}.}
  \label{fig:final_spectrum}
\end{figure}

To demonstrate the final result, Fig.~\ref{fig:final_spectrum} shows the one-dimensional spectrum of NGC~4151 obtained after reduction with \texttt{ARXSP} and subsequent processing in \texttt{IRAF}. The \texttt{IRAF} stage applies the standard long-slit reduction sequence-wavelength calibration, rectification of the spectral trace and extraction of the one-dimensional spectrum-performed according to established methodology (Serebryansky et al. 2025). The data were plotted in \texttt{Python} and illustrate the full workflow from digitized images to calibrated spectra ready for scientific analysis.

During the testing of \texttt{ARXSP}, more than 500 spectra of active galactic nuclei from the archive of the Fesenkov Astrophysical Institute were processed. Each spectrum was accompanied by a set of 3--7 calibration frames. The final data are presented in a standardized format and have been published on the KazVO service\footnote{\href{https://vo.fai.kz/obs_data.php?path=/spectra_agn_archive/q/web/form}{https://vo.fai.kz/obs\_data.php?path=/spectra\_agn\_archive/q/web/form}}, thereby ensuring their availability for integration into modern astrophysical research and VO tools.

\section{Conclusion} \label{sec:conclusion}

This paper has presented the modular desktop application \texttt{ARXSP}, designed for the reduction of archival spectral data and for addressing the specific challenges of processing digitized astronomical plates. The program integrates specialized modules for geometric correction, removal of S-shaped distortions, construction of the characteristic curve, and conversion of values into relative intensities.  

The key advantages of \texttt{ARXSP} include:
\begin{itemize}
    \item support for the full preprocessing cycle of EOC spectra, from geometric correction to the transformation of optical density into intensity;
    \item a modular architecture with automatic integration of new components;
    \item reproducibility of processing ensured through automatic logging of all modifications in the FITS headers;
    \item cross-platform compatibility with \texttt{Windows}, \texttt{macOS}, and \texttt{Linux} without the need for complex environment configuration.
\end{itemize}

At the same time, the current version of the program has several limitations. In particular, automated batch processing of large datasets is not yet implemented; support for modern CCD data is restricted to basic operations, since the main focus has been on the preprocessing of digitized spectra; further analysis must be carried out using external packages such as \texttt{IRAF}.  

Future development is planned to expand the functionality of individual modules, in particular by extending preprocessing methods for data obtained with modern CCD detectors. The integration of elementary stages of further reduction, including the transformation of spectra into the “wavelength–intensity” scale within a unified interface, is also under consideration.  

In this context, \texttt{ARXSP} not only achieves its primary objectives but also contributes to the long-term preservation and scientific reuse of historical spectral archives. By providing digitized data in a standardized format, the application facilitates their integration into international Virtual Observatory frameworks and ensures their availability for modern astrophysical research.

\section{Acknowledgment}

This research is funded by the Committee of Science of the Ministry of Science and Higher Education of the Republic of Kazakhstan (Grant No.~AP22784884).

The authors would like to express their sincere gratitude to Dr. L.~N.~Kondratyeva and Dr. E.~K.~Denissyuk for developing the general data reduction algorithm on which our program is based.  

Special thanks to Dr. L.~N.~Kondratyeva for her numerous consultations and support during the development process.  

We also thank Dr. V.~Kim for valuable advice on spectral data processing, and D.~Anarbek for assistance and consultations in creating installation packages for the software.

\bibliographystyle{elsarticle-harv}
\bibliography{references}

\end{document}